\documentclass[12pt,preprint]{aastex}

\def\plotfiddle#1#2#3#4#5#6#7{\centering \leavevmode
\vbox to#2{\rule{0pt}{#2}}
\includegraphics{#1}}

\begin{document}
\title{Interferometer Observations of Subparsec-scale Infrared Emission 
in the Nucleus of NGC 4151}

\author {
M.~Swain\altaffilmark{1},
G.~Vasisht\altaffilmark{1},
R.~Akeson\altaffilmark{2}, 
J.~Monnier\altaffilmark{3},
R.~Millan-Gabet\altaffilmark{2},
E.~Serabyn\altaffilmark{1},
M.~Creech-Eakman\altaffilmark{1},
G.~van Belle\altaffilmark{2},
J.~Beletic\altaffilmark{4},
C.~Beichman\altaffilmark{1},
A.~Boden\altaffilmark{2},
A.~Booth\altaffilmark{1},
M.~Colavita\altaffilmark{1},
J.~Gathright\altaffilmark{4},
M.~Hrynevych\altaffilmark{4},
C.~Koresko\altaffilmark{2},
D.~Le Mignant\altaffilmark{4},
R.~Ligon\altaffilmark{1},
B.~Mennesson\altaffilmark{1},
C.~Neyman\altaffilmark{4},
A.~Sargent\altaffilmark{2},
M.~Shao\altaffilmark{1},
R.~Thompson\altaffilmark{1},
S.~Unwin\altaffilmark{1},
P.~Wizinowich\altaffilmark{4}
}

\altaffiltext{1}
{Jet Propulsion Laboratory, California Institute of Technology, 
4800 Oak Grove Dr., Pasadena, CA 91109}
\altaffiltext{2}
{Michelson Science Center, California Institute of Technology, 
770 S. Wilson Ave., Pasadena, CA 91125}
\altaffiltext{3}{Univ. of Michigan, 941 Dennison Bldg, Ann Arbor, MI, 48109}
\altaffiltext{4}{W. M. Keck Observatory, California Association for 
Research in Astronomy, 65-1120 Mamalahoa Hwy., Kamuela, HI 96743}

\begin{abstract}
  
We report novel, high-angular resolution interferometric measurements
that imply the near-infrared nuclear emission in NGC 4151 is
unexpectedly compact. We have observed the nucleus of NGC 4151 at 2.2
$\mu$m using the two 10-meter Keck telescopes as an interferometer and
find a marginally resolved source $\leq0.1$ pc in diameter.  Our
measurements rule out models in which a majority of the $K$ band
nuclear emission is produced on scales larger than this size.  The
interpretation of our measurement most consistent with other
observations is that the emission mainly originates directly in the
central accretion disk.  This implies that AGN unification models
invoking hot, optically thick dust may not be applicable to NGC 4151.

\end{abstract}

\section{Introduction}

NGC 4151 is the nearest and brightest Seyfert and it has been
identified as the ``archetype'' of Seyfert 1 galaxies \citep{ngs90}.
It has also been called ``one of the most enigmatic'' of galaxies
\citep{cr96} because the origin of the near-infrared nuclear emission
component, which dominates galactic near-infrared emission, is
unclear.  The extent to which the observed near-infrared emission
comes directly from the accretion disk or jet rather than from clouds
of gas or dust that {\it reprocess} energy from the accretion disk is
critical to models that attempt to unify various observational classes
of AGN \citep{a93}.  Over the past 30 years, arguments for both
thermal and nonthermal (synchrotron) emission mechanisms have been
proposed (see \citet{ngs90} and references therein).  Evidence of $K$
band emission variability on 1--2 month timescales
\citep{pps74,rl81,olt99} and weak (0.1--0.5\%) degrees of $H$ and $K$
band linear polarization \citep{krl77,ryp03} can be explained by both
thermal and nonthermal models.  Currently, the prevailing view is that
the near-infrared emission is thermal and arises from the inner edge
of a dust torus that is heated primarily by UV and soft x-ray emission
from the corona of the accretion disk surrounding a massive black hole
\citep{a93}.  In this picture, the source of the near-infrared
emission would appear spatially unresolved by the largest single
aperture infrared telescopes and would have low degrees of linear
polarization.

Observations at 11 $\mu$m \citep{ngs90,rpp03} strongly imply the
presence of a $<$35 pc diameter source where dust grains are heated to
around 200 K and are responsible for the mid-infrared emission.
Estimates for the inner radius of the dust, which could correspond to
the inner edge of a torus, have been as small as about 0.03 pc
\citep{rl81}, with values in the 0.2--1 pc range presently favored
\citep{a93,pk92,etk93,ryp03}.  In some AGN unification schemes, the
torus is supposed to be optically and physically thick \citep{a93}.
The radius of the broad line region (BLR) is about 0.01 pc
\citep{cbb90}, putting the smallest estimates of the dust torus' inner
radius very close to the black hole and accretion disk.

The parent galaxy for the NGC 4151 Seyfert nucleus is nearly face-on
($i=21^{\circ}$ to the line of site) and has a weak radio jet at P.A.=
$77^{\circ}$ E. of N. \citep{mwp03}.  While the radio feature
corresponding to the UV/optical nucleus has been uncertain, the
candidate features contain emission knots that are unresolved on
angular scales of a few milliarcseconds (mas) and have a surface
brightnesses of a few mJy/beam \citep{mwp03,urc98}.  The 2MASS $K$
band flux for NGC 4151 is 8.519 $\pm$ 0.018 magnitudes, and at least
90\% of the flux is contained in the central 1 arcsec
\citep{krl77}. The bolometric luminosity of the AGN is about $10^{44}$
erg s$^{-1}$, and a $10^7$ M$_{\odot}$ black hole is thought to lie at
the center of the BLR \citep{ulr00}.  Following \citet{mwp03}, we have
taken the distance as 13.3 Mpc (for $H_{0}=75$ km s$^{-1}$), where 1
mas corresponds to 0.065 pc (75.3 light days) in projection.

Interferometric combination of the twin Keck telescopes \citep{col02}
provides sensitivity to 2.2 $\mu$m emission on angular scales of
$\sim$ 1 mas.  Our observations represent a ten-fold improvement in
angular resolution when compared to previous near-infrared
measurements of AGN and make it possible to test the subparsec-scale,
near-infrared emission models of NGC 4151.  Optical/IR interferometry
has traditionally been limited to relatively bright stellar objects.
These observations represent the first measurement of an extragalactic
source with an optical/IR interferometer.

\section{Observations and Data Reduction}

\subsection{Observations}

The Keck Interferometer combines the light from the two 10 m Keck
telescopes, separated by 85 m with the baseline oriented 38$^{\circ}$
(E of N).  Both telescopes are equipped with adaptive optics systems
\citep{wiz02}, which are required for $K$ band interferometer
observations.  The interferometer field of view on the sky is $\sim$50
mas in the $K$ band, as set by a single-mode fiber ahead of the
detector; the interferometer is not sensitive to emission outside of
this field of view.  The fringe tracker uses a 4-bin synchronous fringe
demodulation algorithm \citep{vbc02}, similar to that used at the
Palomar Testbed Interferometer \citep{col99b}.  For the data presented
here, the system operated at a 200~Hz frame rate (the 4-bin
acquisition time).

NGC 4151 was observed with the Keck Interferometer on 2003 May 20 and
21.  The data presented here are from the white-light channel
($\lambda_{\rm center}$=2.18~$\mu$m and $\Delta \lambda
\sim$~0.3$\mu$m).  Observations consist of a series of interleaved
integrations on the source and several calibrators.  Each integration
includes 120 seconds of fringe data followed by a background
measurement.  Included with the fringe data are periodic measurements
of the single-telescope fluxes.  The data are presented as the
visibility amplitude squared, normalized such that an unresolved
object has V$^2=1.0$.  On May 20, only limited data were collected on
NGC~4151 and calibrators due to weather; however, the source was
detected and resolved on both nights, with internally consistent
visibility.  In the results discussed below we have used data only from
May 21.

\subsection{Data Reduction}

The system visibility, the instrument response to a point
source, is measured with respect to the calibrator stars
HIP58918, HD 105925, and HIP60286.  The calibrator angular sizes were
derived by fitting photometry from SIMBAD and 2MASS.  All of the
calibrators have estimated angular diameters of less than 0.2 mas and
are unresolved by the interferometer.  The calibrator angular size
uncertainty was set to 0.1 mas.  Source and calibrator data were
corrected for biases using sky calibrations as described by
\citet{col99} and averaged into blocks of 5 seconds each.  The
calibration procedure includes a per-scan correction for the bias due
to mismatched fluxes between the two telescopes (which includes Strehl
mismatch) using the single-telescope flux measurements.

The data were then calibrated for the system visibility \citep{bod98}.
The calibrated data points for the target source are the average of
the 5 seconds blocks in each integration, with an uncertainty given by
the quadrature of the internal scatter and the uncertainty in the
calibrator size.  In addition to the measurement error, we have
estimated any systematic error in the calibrated visibility to be less
than 0.05 \citep{col03}.  This systematic error was summed
quadratically with the measurement and calibration errors.  The
resulting average visibility is 0.84 $\pm$ 0.06 at an average
projected baseline of 82.7 m at a P.A. = 37\arcdeg E of N.  The
calibrated visibilities and the system visibility estimates are shown
in Fig. 1.

\section{Discussion}

\subsection{Compact Emission}

If all the light in the fringe tracker field of view (approximately
the diffraction limit of the single telescope) came from an unresolved
point source, the interferometer calibrated $V^2$ would be unity.  Our
measured $V^2$ is relatively high, implying the source of this
emission is small.  A variety of compact flux distributions could
produce the measured visibility; we have used three simple, geometric
models to estimate the angular size of the emission.  The models
correspond to a ``face-on'' inclination and are constrained to fit the
measured $V^{2}$.  If the emission is distributed as a
single-component Gaussian, the FWHM = 0.98 mas $\pm$ 0.18 mas.  If the
the emission is distributed as a ring, the ring inner diameter = 0.96
mas $\pm$ 0.26 with a width = 0.06 mas, where the width is set to
match the total $K$ band flux for an optically thick medium with a
temperature of 1500 k.  If the emission is distributed as a point
source with an extended (over-resolved) component, the point-source
fraction of the flux = 0.92 $\pm$ 0.04 of the flux.  Whatever the
details of the emission distribution or mechanism, the bulk of the
emission comes from a compact ($\leq$1.52 mas at the $3\sigma$ level )
region.  Our measurements rule out any scenario with a majority of the
$K$ band nuclear emission coming from a region with a radius in
projection larger than $0.05$ pc.

\subsection{Possible Emission Mechanisms}

We consider four possible mechanisms for the $K$ band emission.  Because
of the combined constraints implied by our measurement and other
observations, we find two of the cases to be unlikely.

\noindent {\bf Synchrotron Emission:} VLBA imaging of NGC 4151 at a
few mas scales \citep{urc98} reveals that its parsec-scale radio jet
is subluminous; it is several orders of magnitude fainter than the
parsec-scale radio jets imaged in nearby classical radio galaxies.
Measurements at 18 cm and 6 cm of the nuclear region reveal a
relatively flat spectrum ($\nu^{\alpha}$ with $\alpha \simeq -0.3$)
source with a peak flux density of about 1 mJy mas$^{-2}$. In
contrast, the large near-infrared flux density of $\simeq 10^2$ mJy
mas$^{-2}$ is inconsistent with the radio fluxes and implies that the
nuclear near-infrared source is unrelated to the radio source and, by
inference, unrelated to a subparsec-scale synchrotron jet.

\noindent {\bf Star Cluster:} Explaining the compact $K$ band emission
as stellar would require $\sim 3\times10^{6}$ O stars in a volume of
$\sim0.001$ pc$^3$.  This explanation is unlikely, as it would imply a
density for young stars $\sim100,000$ times higher than in our own
galactic center \citep{ms96,gnz00}.  A stellar component is also
incompatible with the measured infrared intensity fluctuations
\citep{rl81,qsa00}.

\noindent {\bf Thermal Dust:} If the dust is distributed as a
physically thick torus, the inner radius is much smaller than
typically estimated.  The smallest possible scale for thermal dust
emission is set by the dust sublimation radius.  Using the model of
\citet{bar87}, we find dust with a temperature of 1900 K,
corresponding to a sublimation radius of 0.05 pc, is consistent with
our measurement.  Thus, if the infrared emission is from centrally heated
dust, it is consistent with our observations only if the dust is very
hot and the $K$ band emission is localized near the dust sublimation
radius.

\noindent {\bf Thermal Gas:} A simple, face-on, geometrically thin
accretion disk model, with $T\propto r^{-3/4}$ \citep{ss73} and with
an inner temperature of $\sim 3.5\times 10^5$ K (consistent with the
soft x-ray emission \citep{ulr00}), predicts a $K$ band magnitude of
8.4, 95\% of which is inside a radius of $\sim0.01$ pc.  This model
predicts an unresolved point source, consistent with our size
measurement and with approximately the measured near-infrared flux.

Both the thermal dust and thermal gas models are consistent with our
observations.  However, centrally heated dust, which is thought to
reprocess UV/optical radiation from the central engine, produces a
specific signature known as the ``infrared bump'' \citep{bar87,san89}.
There is no evidence for a comparable feature in NGC 4151
\citep{rl81,em86,alo01}.  Also, a dust torus with a well-defined, but
small, inner radius (corresponding to the sublimation radius) would
give similar time delays between fluctuations in the UV/optical flux
and fluctuations in the $H$ and $K$ band flux.  This is inconsistent
with the analysis by \cite{olt99}.  Thus the absence of the ``infrared
bump'' and the presence of the $H$/$K$ time delay difference both
support the interpretation that the bulk of the infrared flux measured
with the interferometer arises from an unresolved accretion disk.

\section{Summary and Conclusions}

We have measured the angular diameter of the $K$ band emission in the
nucleus of NGC 4151.  Our observations rule out any emission mechanism
that requires a large fraction of the nuclear $K$ band emission to be
produced at a radius of greater than 0.05 pc from the black hole.
Taken in the context of other observations, we interpret our
measurement as evidence that a majority of the $K$ band emission comes
from a central accretion disk.  

\acknowledgments

We dedicate these observations to the memory of Jim Kelley, the late
project manager and tireless advocate of the Keck Interferometer.  We
thank Robert Antonucci, Roger Blandford, and Julian Krolik for useful
discussions.  The Keck Interferometer is funded by the National
Aeronautics and Space Administration as part of its Navigator program.
Part of this work was performed at the Jet Propulsion Laboratory,
the California Institute of Technology, and the Michelson Science Center,
under contract with NASA.  Observations presented were obtained at the
W.M. Keck Observatory, which is operated as a scientific partnership
among the California Institute of Technology, the University of
California, and NASA. The authors wish to recognize and acknowledge
the very significant cultural role and reverence that the summit of
Mauna Kea has within the indigenous Hawaiian community.  We are most
fortunate to have the opportunity to conduct observations from this
mountain. The Observatory was made possible by the generous financial
support of the W.M. Keck Foundation.  This work has made use of
software produced by the Michelson Science Center at the California
Institute of Technology.  This work has also made use of the SIMBAD
database, operated at CDS, Strasbourg, France, and the NASA/IPAC
Infrared Science Archive, operated by the JPL under contract with
NASA.


\clearpage

\begin{figure}
\plotfiddle{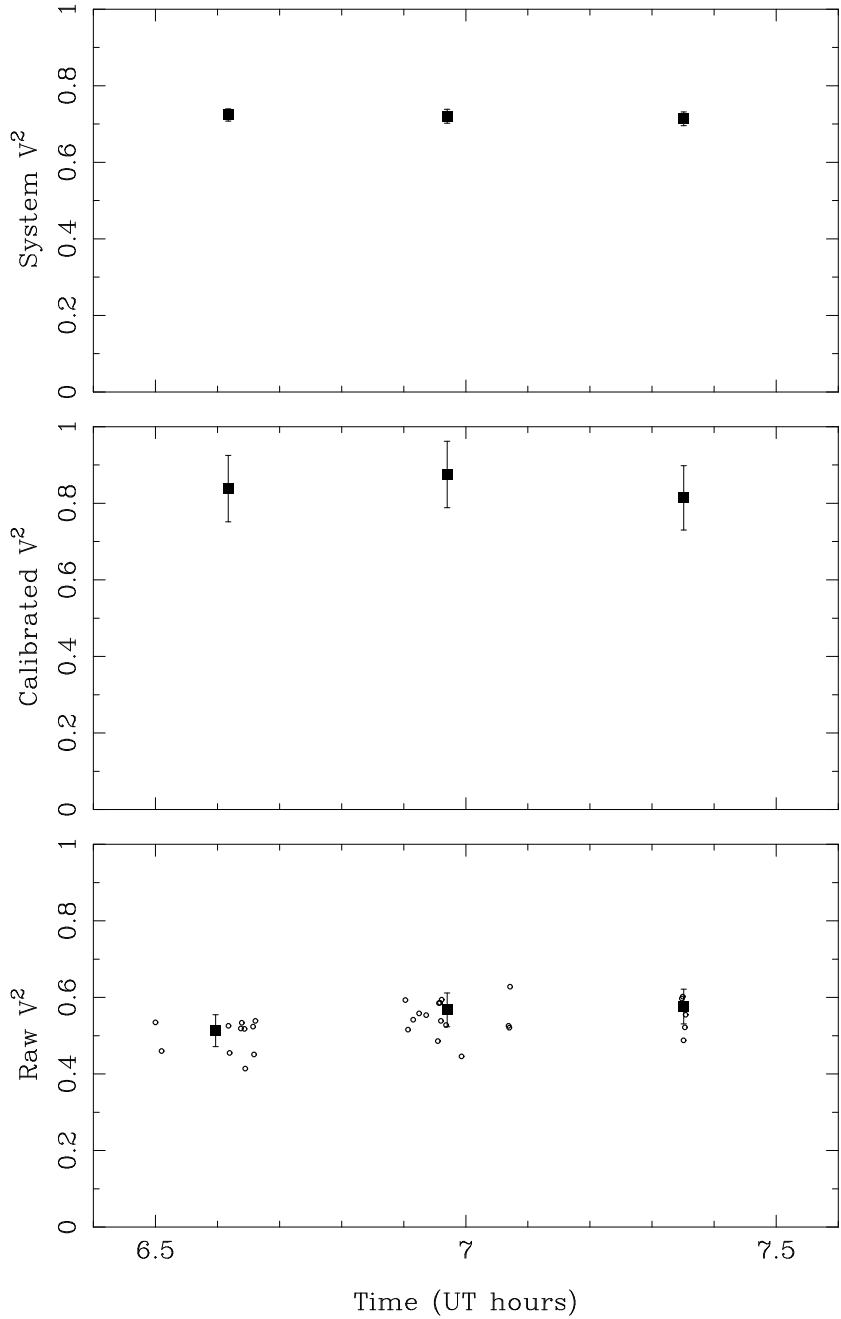}{6.5in}{360}{100}{100}{-300}{-120}
\caption{
The raw (bottom), calibrated (middle) and system visibility (top) for
NGC4151.  In the raw data plot, the 5 sec data averages are shown with
open circles and the scan averages are shown with filled squares.  The
measured visisbility is $0.84\pm0.064$; the error bars in the
calibrated data contain a 5\% systematic component.  The hour angles
for the averaged points are -0.03, 0.33 and 0.71 hours respectively.
The projected baseline length ranges from 81.8 to 83.6 meters and the
position angle from 39.3 to 33.9 degrees (E of N).}
\label{fig:ifvis}
\end{figure} 

\end{document}